\documentclass[reprint,amsmath, amssymb, preprintnumbers, showpacs, showkeys,aps,pra,superscriptaddress,twocolumn]{revtex4-1}

\usepackage{color} 
\usepackage{hyperref}
\usepackage{amsmath,amssymb,mathrsfs}
\usepackage{psfrag}
\usepackage{graphicx}
\usepackage{graphics}
\usepackage{epsfig}
\usepackage{bm}
\usepackage{verbatim,color,ulem}
\usepackage{braket}
\usepackage{verbatim}

\usepackage{tabularx}
\usepackage{tabu}
\usepackage[table]{xcolor}
\definecolor{mygray}{gray}{0.9}
\definecolor{mygray2}{gray}{0.7}
\usepackage{booktabs}
\usepackage{colortbl}
\usepackage{hhline}
\usepackage{transparent}

\newcolumntype{Y}{>{\centering\arraybackslash} X}
\newcolumntype{V}{!{\color{black}\vrule width 1pt}}

\newcommand{\beq}{\begin{equation}}
\newcommand{\eeq}{\end{equation}}

\newcommand{\be}{\begin{equation}}
\newcommand{\ee}{\end{equation}}

\begin{document}

\title{Vortices in self-bound dipolar droplets}

\author{Andr\'e Cidrim}
\affiliation{Departamento de F\'isica, Universidade Federal de S\~ao Carlos, 13565-905 S\~ao Carlos, S\~ao Paulo, Brazil}
\affiliation{Instituto de F\'isica de S\~ao Carlos, Universidade de S\~ao Paulo, C.P. 369, 13560-970 S\~ao Carlos, SP, Brazil}
\author{Francisco E. A. dos Santos}
\affiliation{Departamento de F\'isica, Universidade Federal de S\~ao Carlos, 13565-905 S\~ao Carlos, S\~ao Paulo, Brazil}
\author{Emanuel A. L. Henn}
\affiliation{Instituto de F\'isica de S\~ao Carlos, Universidade de S\~ao Paulo, C.P. 369, 13560-970 S\~ao Carlos, SP, Brazil}
\author{Tommaso Macr\`i}
\affiliation{Departamento de F\'isica Te\'orica e Experimental, Universidade 
Federal do Rio Grande do Norte, and International Institute of Physics, Natal-RN, Brazil}

\begin{abstract}
Quantized vortices have been observed in a variety of superfluid systems, from $^4$He to  
condensates of alkali-metal bosons  and ultracold Fermi gases along the BEC-BCS crossover.
In this article we study the stability of singly quantized vortex lines in dilute dipolar self-bound droplets.
We first discuss the energetic stability region of dipolar vortex excitations within a variational ansatz
in the generalized nonlocal Gross-Pitaevskii functional that includes quantum fluctuation corrections. 
We find a wide region where stationary solutions corresponding to axially-symmetric vortex states exist. 
However, these singly-charged vortex states are shown to be unstable, either by splitting the droplet in two 
fragments or by vortex-line instabilities developed from Kelvin-wave excitations. These observations are the
results of large-scale fully three-dimensional simulations in real time. We conclude with some experimental 
considerations for the observation of such states and suggest possible extensions of this work.
\end{abstract}


\maketitle

\section{Introduction}

Quantized vortices are a direct manifestation of the genuine quantum behavior of superfluid systems.
A prime example is superfluid helium, which has been widely studied over the past decades.
Ultracold quantum gases offer the possibility to investigate vortex properties in a complementary
regime in terms of particle numbers, interaction strength, and range \cite{Dalfovo01,Fetter01}
with either bosonic \cite{Madison00} or fermionic \cite{Zwierlein05} atoms.
Dipolar condensates may also display vortex excitations with peculiar properties
as a consequence of the long range and anisotropy of the interactions \cite{Lahaye09,Martin2017,Abad2009,Tikhonenkov2008}. 
For a three-dimensional condensate with a vortex line and in the presence of a periodic potential, 
the spectrum of transverse modes may display a roton-like minimum~\cite{Wilson08, Shirley14}, which destabilizes 
the straight vortex and leads to a transition from vortex into helical or snakelike configurations \cite{Klawunn08,Klawunn09}.

Theoretical models of superfluid states of ultracold gases at zero temperature are usually based on 
well-established mean-field approximations which accurately describe experiments \cite{Giorgini99},
ranging from analytic treatments and variational approaches to full numerical simulations.
Corrections beyond the mean-field picture have been measured for strongly interacting Bose gases \cite{Navon11}
and for ultracold fermions along the BCS-BEC crossover \cite{Navon10} and compared with {\it ab initio}
quantum Monte Carlo calculations.

The recent observation of ultradilute self-bound droplets both in dipolar condensates \cite{Kadau16,Ferrier16,Chomaz16,Ferrier16-1,Schmitt16,Wenzel17,Chomaz17}, as well
as in two-component Bose mixtures \cite{Cabrera17}, together with a combined theoretical effort, 
established the importance of the fundamental role of quantum fluctuations in ultracold atomic systems 
\cite{Petrov2015,Petrov2016,Baillie16,Wachtler16_fil,Blakie16,Wachtler16,Baillie-exc17,Cinti17,Macia16,Cinti16,Saito16,Boudjemaa17,Laghi17,Cappellaro17,Adhikari17}.
Yet, no work has investigated the presence and stability 
of vortex states in self-bound droplets in ultradilute liquids.

Helium droplets hosting several quantized vortices have been recently observed \cite{Gomez14} 
and studied in detail theoretically \cite{Dalfovo00,Lehmann03,Ancilotto15,Ancilotto17}, 
both in pure samples and in the presence of impurities. The scales are, nevertheless, 
completely different. Helium droplets can be easily taken as a homogeneous, 
infinite superfluid background since vortices are much smaller compared to the 
system size (given the large interaction strengths). 

Droplets in quantum ferrofluids are very anisotropic, and vortex cores are expected to have a size comparable to that of the whole droplet.
Here we address the issue of stability and dynamics of singly quantized 
vortex lines in dipolar droplets for a wide range 
of dipolar interaction strengths and particle numbers. 
We carry out large-scale fully three-dimensional simulations, which allow for an efficient  
determination of energies and shapes of droplets as well as their dynamics. 
We find a strong anisotropy of such droplets, very elongated along the polarization axis.
For small particle numbers, droplets are dynamically unstable towards splitting in 
two droplets where angular momentum
is redistributed into surface collective excitations \cite{Saito02}.
For larger sizes we do not observe splitting, yet vortex lines display bending for long times \cite{Aftalion01,Ripoll01,Bretin03,Simula08,Klawunn09}.
We conclude with a discussion of a possible experimental implementation and 
observation of our findings with current
experimental setups.
\begin{figure*}[t!]
\centering
\includegraphics[width=2.0\columnwidth]{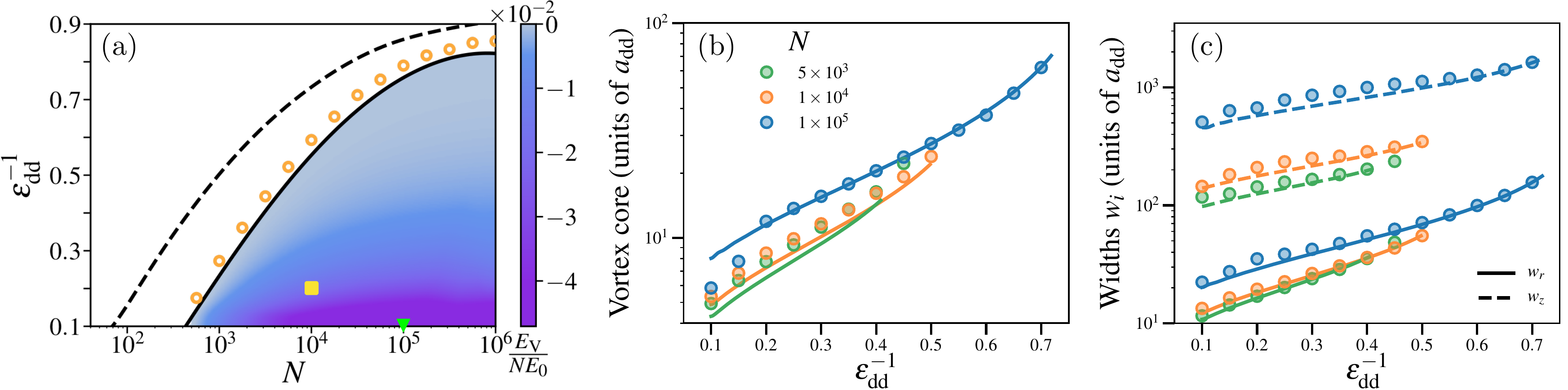}
\caption{(a) Stability diagram of an axisymmetric droplet with a vortex line. 
Density plot of the energy per particle of self-bound solutions hosting a vortex as a 
function of $\varepsilon_{\mathrm{dd}}^{-1}$ and $N$ calculated using the variational Gaussian wave function (\ref{variational}). 
The thick black line corresponds to $E_\mathrm{V} = 0$. 
Energy is in units of $E_0=\hbar^2/(m\,a_\mathrm{dd}^2)$.
The yellow square and the green triangle correspond to 
$(N,\varepsilon_\mathrm{dd}^{-1})=(10^4,0.2)$ and $(10^5,0.1)$ coordinates,
respectively, used in Fig.~\ref{fig2} to characterize the dynamics.
(b),(c) Characterization of droplets hosting a vortex line. 
(b) Size of the vortex core scaled by the dipolar length as a function of 
$\varepsilon_\mathrm{dd}^{-1}$ and for $N=5\times 10^3,10^4,10^5$ particle numbers.
Full dots correspond to energy minima of the Gross-Pitaevskii energy functional.
(c) Longitudinal ($w_z$) and horizontal ($w_r$) droplet size as a function 
of $\varepsilon_\mathrm{dd}^{-1}$ for the same particle numbers as in (b).
In (b) the vortex core size is defined as the distance from the origin to the point where the density reaches 90\% of its maximum. In (c) the radial (axial) width is defined as the mean length $w_{r(z)}\equiv(w_i+w_e)/2$, where $w_i$ and $w_e$ are the distances between the points where the density reaches, respectively, $90 \%$ and $10\%$ of its maximum value along the radial direction (along the vertical, peak-density line).
In all panels lines are variational results, and dots are numerical results for stationary solutions of Eq. (\ref{GPE}), imposing
a vortex with azimuthal symmetry (see text).
}
\label{fig1}
\end{figure*}

\section{Methods}
The dynamics of an untrapped dipolar Bose-Einstein condensate
is described by a generalized non-local Gross-Pitaevskii equation, 
\begin{equation} \label{GPE}
i\, \hbar\, \dot \psi = \left( -\frac{\hbar^2\nabla^2}{2m} + 
\mathcal{K}_\mathrm{int}(\mathbf{r}) + g_\mathrm{LHY}|\psi|^3 \right)\psi,
\end{equation}
where $\psi(\mathbf{r},t)$ is the BEC wavefunction. 
$\mathcal{K}_\mathrm{int}(\mathbf{r})=g_c|\psi(\mathbf{r})|^2+
\int\, \mathrm{d}\mathbf{r'}\,V_\mathrm{dd}(\mathbf{r}-\mathbf{r}')|\psi(\mathbf{r'})|^2$ 
describes the contact and dipolar mean-field interaction of the condensate.  
Here $g_c=4\pi a_s\hbar^2/m$ is the contact interaction strength, with 
$a_s$ being the $s$-wave scattering length, and 
$V_{\mathrm{dd}}(\mathbf{r})=\frac{C_\mathrm{dd}}{4\pi}\frac{1-3\cos^2\theta}{r^3}$ is the dipolar
potential. $C_\mathrm{dd}=\mu_0\mu^2\equiv\frac{12\pi\hbar^2}{m}a_\mathrm{dd}$ is the dipolar coupling constant, $\mu$ is the magnetic dipole moment,  
$\theta$ is the angle between $\mathbf{r}$ and the vertical axis (polarization axis of the dipoles),
and $a_\mathrm{dd}= \frac{\mu_0 \mu^2m}{12\pi\hbar^2}$ is the dipolar length.
The parameter $\varepsilon_{\mathrm{dd}}=a_{\mathrm{dd}}/a_s$ is the ratio of dipolar interaction to the
$s$-wave interaction strengths, defining stability of a uniform condensate in the Bogoliubov approach when
$\varepsilon_\mathrm{dd}<1$ \cite{Lahaye09}.
Quantum fluctuation corrections to the mean-field energy for a 
uniform dipolar condensate are introduced in Eq.(\ref{GPE}) 
by a Lee-Huang-Yang--type (LHY) term, with coefficient
$g_{\mathrm{LHY}}=\frac{128\sqrt{\pi}}{3}\frac{\hbar^2 a_s^{5/2}}{m}\left(1+\frac{3}{2}\varepsilon_{\mathrm{dd}}^2\right)$
\cite{Lima11,Blakie16,Wachtler16_fil}.
Energetic stability of dipolar droplets in trapping potentials have been studied in a number of works. 
A stability diagram for droplets in free space was proposed in \cite{Baillie16} 
via a Gaussian ansatz and checked against numerical simulations. The variational approach 
well describes collective properties of the condensate, such as the energy, the shape close to 
the instability region, and excitation spectra \cite{Wachtler16}. 
\begin{figure*}[t!]
\centering
\includegraphics[width=2.0\columnwidth]{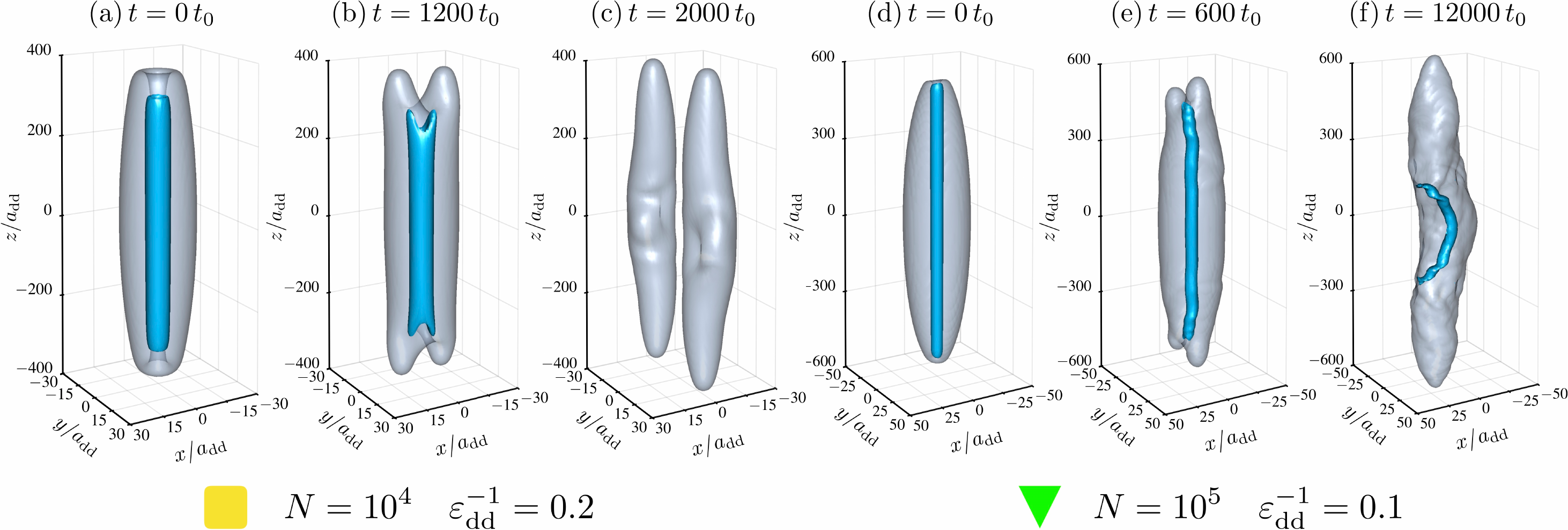}
\caption{ {\it Real-time dynamics of droplets with vortex lines.}
(a), (b), (c) Splitting instability for $N=10^4$ particles and $\varepsilon_\mathrm{dd}^{-1}=0.2$. 
Snapshots of the dynamics at $t/t_0=0,1200,2000$, where $t_0=m\,a_\mathrm{dd}^2/\hbar$.
At intermediate time $t=1200\, t_0$ droplets split in two fragments moving at a finite 
(opposite) momentum and carry excitations with finite angular momentum.
(d), (e), (f) Vortex-line bending for $N=10^5$ and $\varepsilon_\mathrm{dd}^{-1}=0.1$.
Snapshots of the dynamics at $t/t_0=0,600,12\,000$. At initial times unstable Kelvin-wave modes 
tend to split the droplet.
After a transient dynamics, where part of the angular momentum is transferred to surface collective excitations, 
the vortex line bends ($t=12\,000\,t_0$). See the full-dynamics video in the Supplemental Material \cite{supplemental}. 
Notice the difference in the length scales between the axes of the $N=10^4$ and $N=10^5$ plots
due to very strong anisotropy and different droplet lengths. 
Results are obtained by solving numerically Eq. (\ref{GPE}) on a grid size of $512\times512\times256$ points.
}
\label{fig2}
\end{figure*}

\section{Energetic stability diagram}
We begin our study with the static properties of vortex lines introducing a variational wave function
\begin{equation} \label{variational}
\psi_V({\bf r}) = 
\left(\frac{2^{2\ell+3}N}{\pi^\frac{3}{2}\sigma_\rho^{2\ell+2}\sigma_z}\right)^\frac{1}{2}
\rho^\ell\,e^{i\ell\phi}\, e^{-2\left(\frac{\rho^2}{\sigma_\rho^2}+\frac{z^2}{\sigma_z^2} \right)},
\end{equation}
where $N$ is the particle number and $\rho$ the radial coordinate~\cite{Teles13}. The choice $\ell=0$ corresponds to 
a state with no vortex, whereas for $\ell>0$ the state has $\ell$ quanta of circulation. 
In this work we specialize to the case of $\ell=1$, and 
leave the investigation of
multicharged vortices to a separate study. 
In Eq. (\ref{variational}) the widths of the condensate $\sigma_\rho$ and $\sigma_z$ are 
variational parameters to be determined via a minimization of the full energy functional
associated with Eq. (\ref{GPE}).
Therefore, we compute the rescaled energy
\begin{equation} \label{en_vor}
\begin{array}{ccl}
\frac{E_V}{E_0} &=& \frac{1}{N \sigma_z^2}\left(1+ \frac{4}{y^2}\right)+
\frac{2\sqrt{2}}{\pi^\frac{1}{2}\sigma_z^3 y^2 N} \left(\varepsilon_\mathrm{dd}^{-1} - g(y)\right)\\
&&+\frac{8192\sqrt{2}}{625\pi^{5/4}}\frac{1+\frac{3}{2}\varepsilon_\mathrm{dd}^2}{\varepsilon_\mathrm{dd}^{5/2}N^2},
\end{array}
\end{equation}
where $E_0=\frac{\hbar^2}{m\, a_\mathrm{dd}^2}$, $y=\sigma_\rho/\sigma_z$, $g(x)=f(x)+3x\,f'(x)/8+x^2\,f''(x)/8$,
and $f(x)=\frac{1+2x^2}{1-x^2}-\frac{3x^2 \text{arctanh}{\sqrt{1-x^2}}}{(1-x^2)^{3/2}}$ \cite{Lahaye09}.
The resulting minimization of Eq. (\ref{en_vor}) is shown in Fig.~\ref{fig1}(a) as a function of $\varepsilon_\mathrm{dd}^{-1}$ 
and particle number $N$. For comparison we show the result of energy minimization for $\ell=0$ (dashed line) \cite{Baillie16}.
The shaded region below the full line is the energetic stability region of a droplet with $\ell=1$, which is shrunk compared to the $\ell=0$ case.
Above the solid line there is no minimum, and the minimum energy state is a uniform solution 
with vanishing density \cite{footnote1}.

\section{Shape of dipolar droplets with a vortex line}
We proceed by characterizing the shape of the droplet in the presence of a vortex for different particle 
numbers and $\varepsilon_\mathrm{dd}$.
In Fig.~\ref{fig1}(b) we compute the vortex core size in units of $a_\mathrm{dd}$, whereas in Fig.~\ref{fig1}(c) we plot the
horizontal and vertical widths $w_r$ and $w_z$. The vortex core size is defined from the origin over the $z=0$ plane to the point where the density reaches 90\% of its maximum value. The droplet radial (axial) width is defined as the mean length $w_{r(z)}=(w_i+w_e)/2$, where $w_i$ and $w_e$ are the distances between the points where the density reaches, respectively, $90 \%$ and $10\%$ of its maximum value along the radial direction (along the vertical, peak-density line).
Variational calculations are checked against full numerical simulations of Eq. (\ref{GPE}) in imaginary time (dots in Fig.~\ref{fig1}), exploiting
the azimuthal symmetry of vortex states, where we assume $\psi_\mathrm{V}({\bf r})= e^{i\phi}\Psi_\mathrm{V}({\rho},z)$.  
The problem reduces then to an effective two-dimensional (2D) problem for the function $\Psi({\rho},z)$ (solved over a grid size of $512\times 256$ points).
To efficiently compute the kinetic term, we employ a discrete Hankel transform in the radial direction (of first order, to impose the vortex node at $\rho=0$) and the usual fast Fourier transform in the longitudinal direction \cite{footnote2}.
In Figs.~\ref{fig1}(b) and~\ref{fig1}(c) lines terminate at the spinodal point of Fig.~\ref{fig1}(a) for the corresponding value of $N$. 
We notice that for a wide range of interactions and particle numbers the vortex core is of the same order of the radial width $w_r$,
and both are always much smaller than the longitudinal width $w_z$. 
All lengths decrease slightly by increasing $\varepsilon_\mathrm{dd}$ for a fixed $N$. 
For every simulation analyzed in Fig.~\ref{fig1}, both $\rho$ and $z$ domains were 
chosen to be at least twice as large as the maximum extension of the droplet predicted by the 
variational approach $\left(\mathrm{i.e., }\gtrsim 2 \, \sigma_z\right)$. The application of 
an interaction cutoff (as proposed in \cite{Lu2010}) turned out to be irrelevant in these large-domain cases, meaning that the Fourier copies (of the periodically bound numerical box) were sufficiently spaced, making artificial effects of the long-range interaction negligible.
We observe agreement between the variational and numerical results for small $\varepsilon_\mathrm{dd}$ and particle numbers. A discrepancy for large particle numbers and stronger interactions [as is the case of $\varepsilon_\mathrm{dd}^{-1}<0.2$ and $N=10^5$, see Fig.~\ref{fig1}(b)] is expected due to the combined effect of the limited validity of the variational ansatz as well as for resolution issues related to the strong anisotropy
of the droplet.

\renewcommand{\arraystretch}{1.5}

\begin{figure*}[t!]
\centering
\includegraphics[width=2.0\columnwidth]{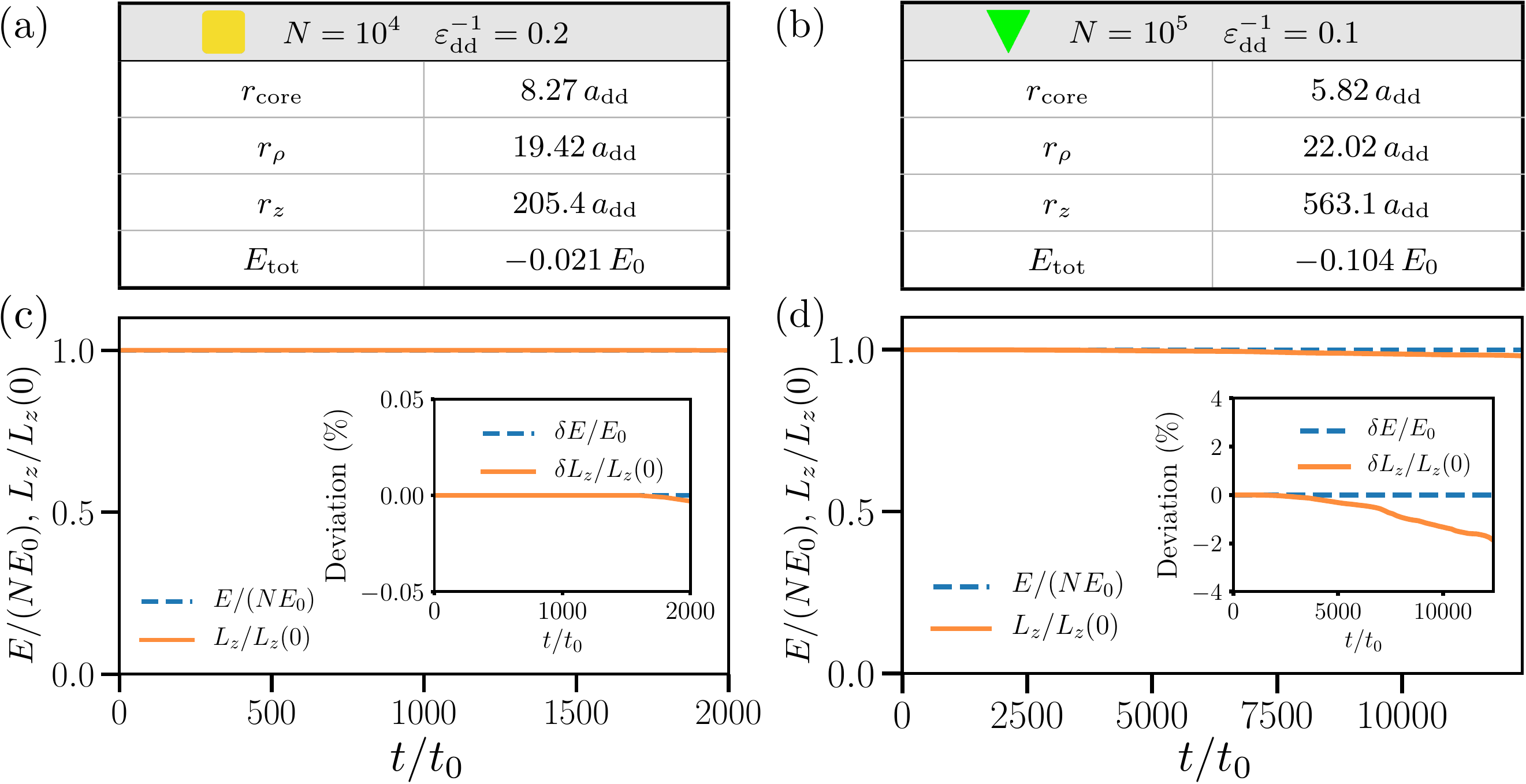}
\caption{{\it Numerical simulations.} Tables (a) and (b) show the vortex core $r_\text{core}$, the longitudinal $r_z$, and the transverse radius $r_\rho$ in units of $a_\text{dd}$ and the total energy per particle in units of $E_\text{0}$ for the two points in the diagram of Fig.~\ref{fig1}. They correspond to the stationary state for (a) $N=10^4$ and $\varepsilon_\mathrm{dd}^{-1}=0.2$, and (b) $N=10^5$ and $\varepsilon_\mathrm{dd}^{-1}=0.1$. Results are obtained by solving numerically Eq. (\ref{GPE}) with imaginary-time evolution on a grid as large as $512\times256$ points along the transverse and longitudinal directions, respectively. The domain ranges are (a) $\rho\times z \in [-300,300]\times [-800,800]$ and (b)  $\rho\times z \in [-400,400]\times [-1000,1000]$ in units of $a_\mathrm{dd}$. Plots (c) and (d) show the energy and longitudinal angular momentum conservation along the dynamics of Fig.~\ref{fig2} for the same particle numbers and interactions as in (a) and (b), respectively.}
\label{fig3}
\end{figure*} 

\section{Real-time dynamics}
A crucial issue, relevant for the experiments, is the stability of such vortex states in self-bound droplets. 
To address this point we perform fully three-dimensional simulations in real time
to take into account possible instabilities which can break azimuthal symmetry (see below). 
The input states at $t=0$ are created from the results of 2D imaginary-time relaxation described previously. These 2D solutions are interpolated to a grid of $512\times512\times256$ points, thus generating three-dimensional initial states [see Figs.~\ref{fig2}(a) and ~\ref{fig2}(d)], slightly perturbed with numerical noise.

In Fig.~\ref{fig2} we illustrate the prototypical real-time dynamics of a droplet with a vortex line at $t=0$ 
for two cases of condensates
of $N=10^4$ and $N=10^5$ particles with $\varepsilon_\mathrm{dd}^{-1} = 0.2$ and $0.1$, respectively.
Time is measured in units of $t_0=m\,a_\mathrm{dd}^2/\hbar$, which equals 
$0.12\,\mu$s for $^{164}$Dy and $0.03\,\mu$s for $^{168}$Er. The blue shaded region is an isosurface cut of the magnitude of the pseudovorticity vector ${\bm \omega} = \bm \nabla \mathrm{Re}\left(\psi\right)\times\,
\bm \nabla \mathrm{Im}\left(\psi\right)$. This quantity is tangent to the vortex line along its length and can be used to track the numerical points corresponding to the vortex core, where the zeros of the imaginary and real parts of the wave function intersect \cite{Serafini17}.
The two dynamics display very different features.
For $N=10^4$ the system develops a splitting instability and divides in two droplets with 
$N_\mathrm{f} \approx N/2$ at $t\approx 1200\,t_0$, 
where $N_\mathrm{f}$ is the particle number of each droplet after splitting. 
For longer times the two fragments move apart with opposite momenta
and display no residual vorticity. All initial angular momentum gets transferred into collective surface excitations
of the droplets.
For the larger system, splitting instability starts to develop at the same time $t=600\,t_0$; however, the droplet
does not fragment. At later times the condensate restores a droplet-like configuration with the 
development of Kelvin waves
along the vortex line, eventually leading to vortex bending for $t=12000\,t_0$ and surface excitations. 
We observe similar features, i.e., absence of splitting and enhancement of surface excitations, 
also for longer times. The shape of the bent vortex line resembles the 
$U$-shaped excitations studied in \cite{Aftalion2003}, which were shown to hold less angular momentum 
than a straight-line vortex state, corroborating our description of angular momentum being transferred to 
surface modes [see Fig.~\ref{fig2} (f)]. 
Since external torques on the droplet are absent, no change of the angular momentum along any direction is
expected. Its necessary conservation means that angular momentum had to be transferred from the vortex line to surface modes. Numerically, a finer grid and larger simulation box would be needed in order to preserve the angular momentum along the vertical direction for very long times.
In Figs.~\ref{fig3} (c) and~\ref{fig3}(d) we show the numerical deviation of the energy and longitudinal angular momentum for the 
two cases discussed above.

\section{Experimental considerations}
Vortices in ultracold atomic systems are controllably created either by phase imprinting or via an effective rotating 
potential generated by an applied laser beam. In any case, vortex imprint must be done in-trap, 
following the droplet preparation and before its release. The “optical spoon” technique is more invasive 
and requires longer equilibration times and it becomes less likely to work experimentally. Phase imprinting, 
on its turn, can be done with high spatial resolution as well as by means of very short optical pulses, 
allowing an almost instantaneous imprint of a vortex. Following the simulations presented above, the 
typical times for the observation of vortex dynamics for $^{164}$Dy range from $100\,\mu$s to just below $2$ ms, 
which is short but still within experimental resolution. 
For example, for the parameters as in Fig.~\ref{fig2}(a), splitting instability sets on for times 
$t\approx 10^3\ t_0$, which correspond to $t\approx 0.1$ ms for $^{164}$Dy. Nevertheless, we have also verified instances in which the onset of the splitting process took slightly longer time. This was the case of $\varepsilon_\mathrm{dd}^{-1}= 0.6$ and $N=10^5$, where $t \approx 45\,000\ t_0$, corresponding to $t \approx 6$ ms. Detection, in any case, must be done \textit{in situ}, 
preferably with a levitating external magnetic field gradient to allow for longer observation times  \cite{Wenzel17}. 
Extraction of dynamical properties, e.g., momentum, shape, atom-number, and/or absence of droplet movement in the cases covered in Fig.~\ref{fig2}, can be done at longer evolution times, on the order of several milliseconds, when detection is expected to be easier.

\section{Conclusions}
In this article we studied the stability of quantum vortex lines in dilute self-bound droplets of dipolar atoms. 
We first discussed the energetic stability region of such vortex excitations via a variational ansatz  
in the generalized nonlocal Gross-Pitaevskii functional that includes a LHY-type contribution. 
The region corresponding to the stationary solutions where $E_\mathrm{V}<0$ is unstable to 
fragmentation into two droplets. 
When this is not the case we found that Kelvin waves establish along the vortex line, which eventually bends
in the central region of the droplet. We confirmed our findings by detailed fully three-dimensional 
numerical simulations of vortex states created by phase imprinting.
The situation where Kelvin waves start developing has also been predicted to appear in a similar context 
of three-dimensional dipolar BECs \cite{Klawunn08}. Droplets with vortices may thus serve as promising test-beds to 
the study of twisted vortex lines in real-life experiments.
An extension of this work would include the investigation of the excitation
spectrum of these vortex lines, similarly to what has been recently done in vortex-free droplets \cite{Baillie-exc17},
and with vortex states in trapped geometries \cite{Wilson09}.
These instabilities offer new opportunities for devising stabilization methods, such as temporal or spatial
modulation of the scattering length as proposed for nondipolar BECs \cite{Malomed03,Adhikari04} or pinning potentials \cite{Isoshima99,Kuopanportti10}.
Also, the appearance of vortex arrays \cite{Cooper05,Zhang05} as well as the effects of
impurities and turbulence phenomena may be relevant to current experiments \cite{Barenghi14,Tsatsos16,Tsubota17,Bland17}.

\begin{acknowledgments}

We thank V. S. Bagnato for valuable discussions and A. Cappellaro
for carefully reading the manuscript. 
Support from CePOF through FAPESP 2013/07276-1 is acknowledged.
T.M. acknowledges CNPq for support through Bolsa de Produtividade 
em Pesquisa No. 311079/2015-6.
E.A.L.H. acknowledges support from FAPESP through Grant No. 2015/20475-9. F.E.A.S. acknowledges CNPq for support through Bolsa de Produtividade em Pesquisa No. 305586/2017-3. A.C. is supported by FAPESP through Grant No. 2017/09390-7. This research was developed with the help of {\footnotesize XMDS}2 software \cite{Dennis2013}, making use of the computational resources of the Center for Mathematical Sciences Applied to Industry (CeMEAI) financed by FAPESP. 
\end{acknowledgments}

\bibliographystyle{unsrt}

\begin{thebibliography}{99}
\bibitem{footnote1} We checked from the variational ansatz 
that metastable solutions for which a local minimum has positive energy
exist and are within the width of the black solid line.
\end{thebibliography}

\end{document}